\documentclass{aa}  
\usepackage{graphicx}
\usepackage[varg]{txfonts}
\usepackage{color}
\usepackage{natbib}
\bibpunct{(}{)}{;}{a}{}{,}
\usepackage{amstext}
\usepackage[normalem]{ulem}

\begin{document}

\title{Where are compact groups in the local Universe?}
\author{Eugenia D\'iaz-Gim\'enez\thanks{eugeniadiazz@gmail.com}   
\and Ariel Zandivarez} 

\institute{
Instituto de Astronom\'{\i}a Te\'orica y Experimental, IATE, CONICET, C\'ordoba, Argentina.\\
Observatorio Astron\'omico, Universidad Nacional de C\'ordoba, Laprida 854, X5000BGR, 
C\'ordoba, Argentina
}
\date{Received XXX; accepted XXX}
\abstract{}
{The purpose of this work is to perform a statistical analysis of the location of
compact groups in the Universe from observational and semi-analytical points of view.
} 
{We used the velocity-filtered compact group sample extracted from the Two Micron
All Sky Survey for our analysis. We also used a new sample of galaxy groups 
identified in the 2M++ galaxy redshift catalogue as tracers of the large-scale structure.
We defined a procedure to search in redshift space for compact groups that can be
considered embedded in other overdense systems and applied this criterion to
several possible combinations of different compact and galaxy group subsamples.
We also performed similar analyses for simulated compact and galaxy
groups identified in a 2M++ mock galaxy catalogue constructed from the Millennium Run
Simulation I plus a semi-analytical model of galaxy formation.
}
{We observed that only 
$\sim27\%$ of the compact groups can be considered to be embedded in larger
overdense systems, that is, most of the compact groups are more likely to be
 isolated systems. 
The embedded compact groups show statistically smaller sizes and 
brighter surface brightnesses than non-embedded systems.
No evidence was found that embedded compact groups are more likely to inhabit 
galaxy groups with a given virial mass or with a particular dynamical state.
We found very similar results when the analysis
was performed using mock compact and galaxy groups.
Based on the semi-analytical studies, we predict 
that 70\% of the embedded compact groups probably are 3D physically dense systems.
Finally, real space information allowed us to
reveal the bimodal behaviour of the distribution of 3D minimum distances 
between compact and galaxy groups.  
}
{
The location of compact groups should be carefully taken into account when comparing 
properties of galaxies in \emph{\textup{environments that are
a priori}} different
\thanks{Tables~\ref{at1} and~\ref{at2} are only available in electronic
form at the CDS via anonymous ftp to 
cdsarc.u-strasbg.fr (130.79.128.5) or via http://cdsweb.u-strasbg.fr/cgi-bin/qcat?J/A+A/ }
.}

\keywords{Methods: numerical -- Methods: statistical -- Galaxies: groups: general}
\titlerunning{Compact groups and the large scale structure}
\maketitle
\section{Introduction} 
\label{intro}
The search for clues that help in unveiling the formation scenario
of different structures in the Universe 
is one of the main goals in current extragalactic astronomy.
These clues are searched for everywhere to understand the formation 
of galaxy systems and the evolution of galaxies in these systems. 
Studies conducted to 
improve our understanding of galaxy evolution focus on compact
groups because of their extreme nature.
\begin{figure*}
\begin{center}
\includegraphics[width=\hsize]{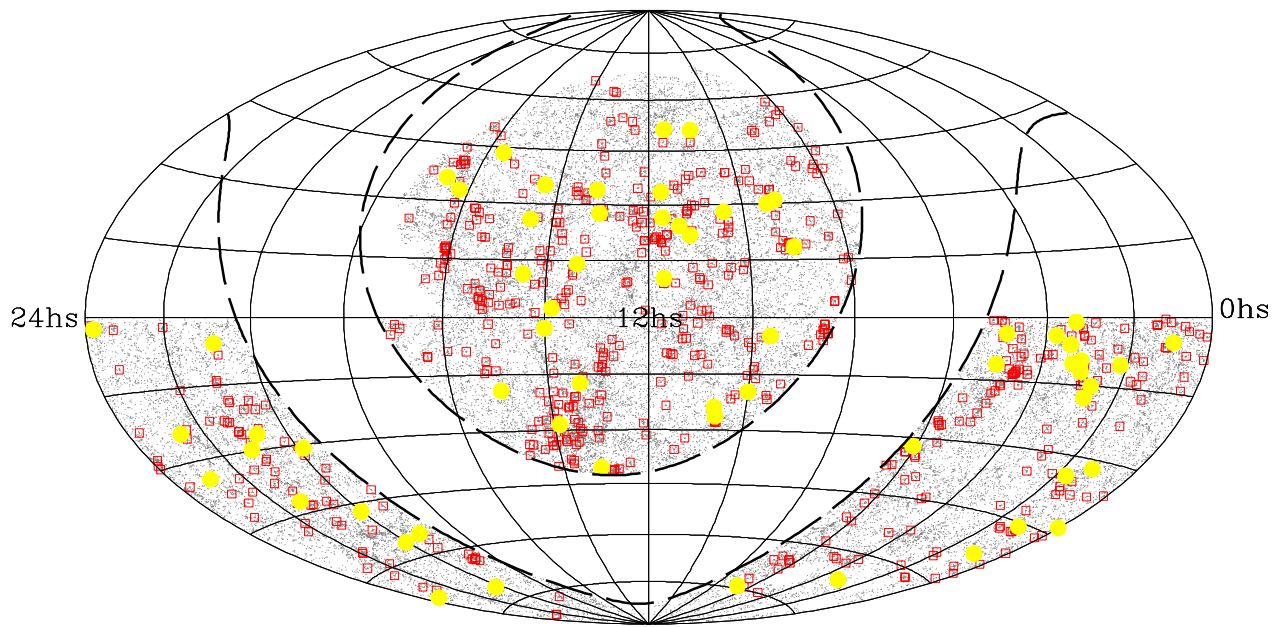}
\caption{Aitoff projection of galaxies in the 2M++ with a K-apparent magnitude lower than
12.5 (grey points) and excluding the region around the Galactic Plane
(dashed lines). \emph{\textup{Open squares}} represent the 583 galaxy groups identified in 
the 2M++ catalogue using a contour overdensity contrast of 433 with a group radial velocity
lower than $12500 \, \rm km \, s^{-1}$, and excluding galaxy groups that are closely related to CGs,
while \emph{\textup{filled circles}} are the 63 2MCGs that lie in the restricted
2M++ area used in this work 
}
\label{f1}
\end{center}
\end{figure*}

The discovery of compact groups began with \cite{Stephan1877} and 
\cite{Seyfert48}. Later, the
 initial discovery of a very compact association of red objects by Shakbazyan in 1957
 was confirmed as a galaxy compact group by \cite{robinson73}. 
This study triggered a systematic search 
for this type of system. 
The pioneer works on Shakbazyan compact groups 
\citep{shak73,shak74,baier74,petrosian74,petrosian78,baier75,baier76a,baier76b,baier78,baier79}
were followed by the attempt of \cite{rose77} to construct a compact group catalogue,
which in turn was followed by the most widely known and most frequently studied Hickson Compact Group catalogue 
(HCG, \citealt{hickson82,hickson97} and references therein).
These small systems of a few galaxies in close proximity have
caused the construction of several catalogues of compact groups 
(e.g. \citealt{prandoni94,Barton+96,FK02,Lee+04,mcconnachie09,diaz12})
as well as a host of scientific analyses of their physical 
properties. Many of these studies have focused on 
their internal structure by analysing the main 
characteristics of the galaxy members and aiming to distinguish the nature
of compact groups as physical entities (e.g. \citealt{mamon86,
mendes91,moles94,olmo95,verdes97,verdes01,verdes05,kelm04,martinez10,plauchu12}).
Studies were also conducted with the aim of understanding the 
different formation scenarios that might lead to their
small projected sizes (e.g. \citealt{hickrood88,hernquist95,tov01}).
There are also several studies trying to differentiate
between the properties of compact group galaxies and galaxies inhabiting 
different environments such as loose groups or in the field 
(e.g. \citealt{krusch03,decarvalho03,proctor04,delarosa07,coenda12,mart13}).

To understand the real nature of compact groups,
it is important to study the environment that these 
galaxy systems inhabit. 
A wider picture of the formation scenario of compact groups 
can be obtained when their surroundings are taken into account.    
Several attempts have been made in the past to solve this question.
For instance, \cite{rood94} and \cite{barton98}
stated that compact groups are not fully isolated
systems. They found instead that between 50\% and 70\% of compact groups are
embedded in overdense regions such as loose groups or clusters of galaxies. 
However, \cite{palumbo95} stated that only $\sim 20\%$ of compact groups
are close to extended concentrations of galaxies.
Subsequent studies also reflected differences in the percentages of 
compact groups that can be associated with larger structures, ranging from $\sim 30\%$ to 50\% \citep{andernach05,decarvalho05}.
Similar results were obtained recently by \cite{mendel11} using
a large sample of compact groups that was
identified in the Sloan Digital Sky Survey 
Data Release Six \citep{dr6} by \cite{mcconnachie09}. \cite{mendel11} also found that 
different galaxy populations are observed
in the neighbourhood of embedded or isolated compact groups. 
However, the compact group sample used in the
later work lacks fully spectroscopic information for each galaxy member
in the systems, and the compact groups in the sample are not homogeneous 
in the definition of their galaxy members, since the chance to find all the group 
members depends on the magnitude of the brightest 
galaxy of the group.
Given the magnitude limit of the parent catalogue, it is not possible to establish for more than 85\% of the groups in this
sample whether they fully meet the compact group criteria, 
and only around 20 systems have been spectroscopically confirmed as concordant 
compact groups from their radial velocities.
Hence, it is important to support the studies of the distribution of compact groups
using samples with well-known and homogeneous selection criteria  
to improve our understanding of compact groups 
and the effect that their location 
in the large-scale structure of the Universe might have on  
their formation history and on the properties of their galaxies.

Recently, one of the largest samples of compact groups
with confirmed membership using spectroscopic information
has been extracted from the Two Micron All Sky Survey (2MASS) \citep{2mass}.
\cite{diaz12} have produced a very
reliable sample of compact groups with several advantages:
i) it is the largest available sample of velocity-filtered compact groups
in the local Universe with at least four members of similar luminosity;
ii) this sample has restricted the apparent magnitude of the 
brightest galaxy in the groups to ensure that all
members can 
span a range of three magnitudes;
iii) it is selected by stellar mass (K band), which is expected
to be a better tracer for magnitude gaps and luminosity
segregation; and iv) this sample shows statistical signs of
mergers; mergers are expected in physically dense groups.
Therefore, this sample of compact groups gives us a very suitable
opportunity to revisit the evidence about the location of these
systems in the large-scale structure of the local Universe.

On the other hand, the statistical analyses performed 
for these peculiar systems using numerical simulations plus semi-analytic 
models of galaxy formation have been of great help in the last years. 
Among them are \cite{mcconnachie08}
and \cite{diaz10}, who determined that between 50-70\% of compact
groups identified using Hickson's criterion can be considered
physically dense groups. They also demonstrated the incompleteness
of the Hickson sample.
Recently, and using similar semi-analytical 
tools, a new analysis of the influence of these extreme
environments on their faint galaxy population has been
performed by \cite{faint14}. These authors reported that when compared with suitable control group samples, 
the compact group environment is not hostile enough to
modify the distribution and density of the faint galaxy 
population that resides in the group and in their surroundings. 
Moreover, the authors also compared their results with those 
obtained using the observational sample of compact groups 
of \cite{diaz12}, which statistically confirmed their semi-analytical 
findings. 

Therefore, the aim of this work is twofold: using the compact group
sample extracted by \cite{diaz12} from the 2MASS catalogue, we analyse the
location of these systems in the underlying large scale structure
traced by a new galaxy group sample extracted from the 2M++ catalogue; and 
we analyse whether the semi-analytical systems (compact and galaxy groups) 
extracted from a mock catalogue show a spatial distribution that resembles 
the results obtained from observations.

This paper is organised as follows: 
in Sect. 2 we describe the observational samples, that is, the catalogues and different procedures adopted to identify the compact groups
and the galaxy groups.
In Sect. 3 we determine the fraction of compact groups that can be considered as embedded systems,
while in Sect. 4 we perform a similar
analysis, but using compact and galaxy groups extracted from mock
catalogues constructed from N-body simulations plus a semi-analytic model 
of galaxy formation. Finally, in Sect. 5 we summarise our results.

\section{Samples}
\begin{figure}
\begin{center}
\includegraphics[width=\hsize]{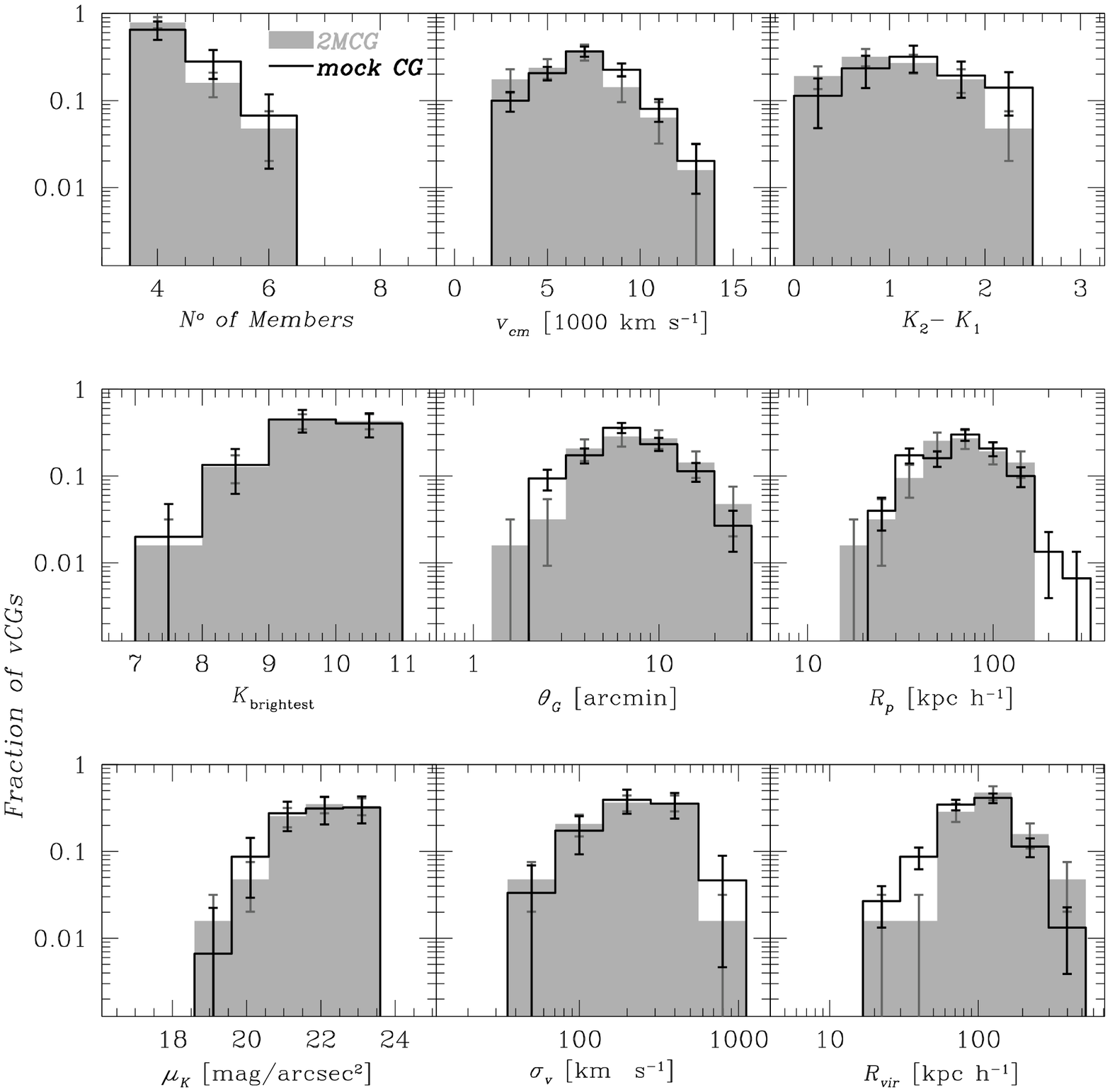}
\vskip -0.05cm
\includegraphics[width=\hsize]{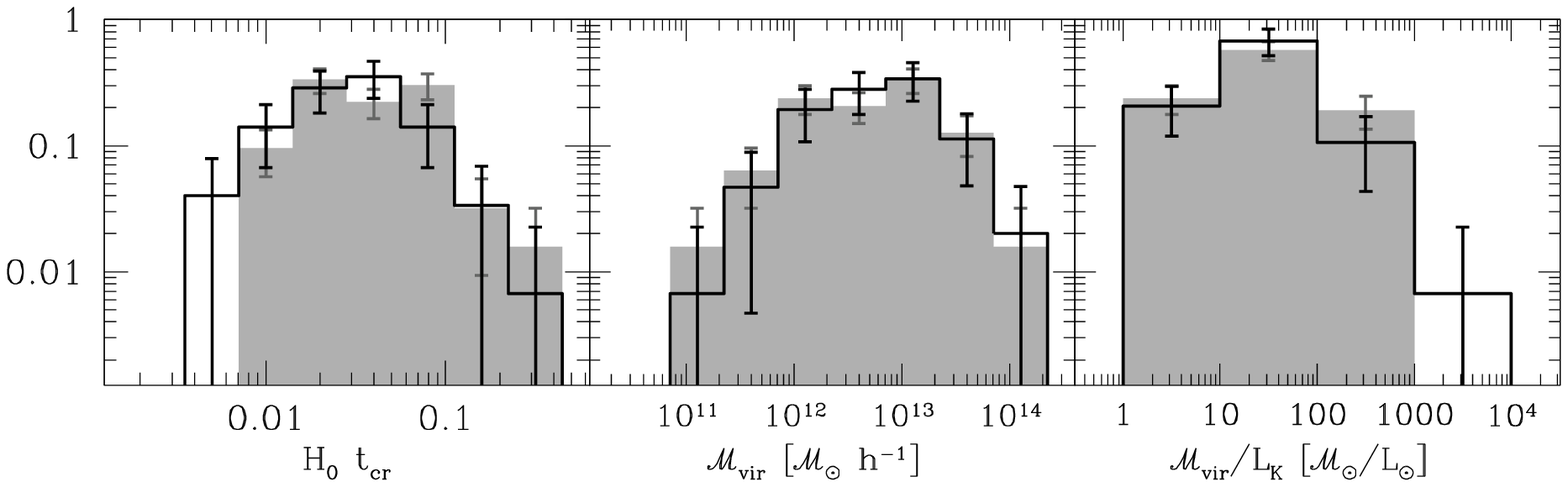}
\caption{ 
Distributions of observable properties of observational and semi-analytical CGs. 
Number of members (top left panel), median 
radial velocity (top centre panel), 
difference in absolute magnitude between the brightest and the 
second brightest galaxies (top right panel),
$K_s$-band apparent magnitude of the brightest galaxy member (second row, left panel),
angular diameter of the smallest circle that encloses the galaxy members 
(second row, centre panel), 
projected group radius (second row, right panel),
group surface brightness (third row, left panel),
radial velocity dispersion (third row, centre panel), 
group virial radius (third row, right panel),
dimensionless crossing time (bottom left panel),
group virial mass (bottom centre panel),
and mass-to-light ratio (bottom right panel).
Grey shaded histograms correspond to the observational sample of CGs (2MCGs) 
that lie on the 2M++ restricted area, 
while black empty histograms correspond to semi-analytical CGs. 
Error bars correspond to Poisson errors.}
\label{f2}
\end{center}
\end{figure}
\subsection{Compact group sample}
\label{compact}
We used the compact group sample (CGs) identified by \cite{diaz12} 
in the 2MASS catalogue avoiding the Galactic plane (defined by Galactic latitude $|b|\le 20$). 
In this section, we briefly describe the procedure and the results found in that work.

The automated searching algorithm used in \citeauthor{diaz12} mimics the procedure defined 
by \cite{hickson82}, and their compact groups satisfy the criteria
\begin{itemize}
\item $4 \le N \le 10$ (population),
\item $\mu_K \le 23.6 \, \rm mag \, arcsec^{-2}$ (compactness),
\item $\Theta_N > 3\Theta_G$ (isolation),
\item $K_{brightest} \le  K_{lim}-3=10.57$ (flux limit),
\end{itemize}
where N is the number of galaxies whose K-band magnitudes
satisfy $K < K_{brightest} + 3$, and $K_{brightest}$ is the apparent magnitude
of the brightest galaxy of the group; $\mu_K$ is the mean K-band surface
brightness, averaged over the smallest circle circumscribing
the galaxy centres; $\Theta_G$ is the angular diameter of the smallest circumscribed
circle; $\Theta_N$ is the angular diameter of the largest
concentric circle that contains no other galaxies within the considered
magnitude range or brighter. Using this algorithm and visual inspection, 
they found 230 CGs in the 2MASS XSC.

They also introduced a filtering  of CGs according to velocity, that is, selecting CGs with
$|v_i -\langle v \rangle| \le 1000 \ \rm km/s $, where $v_i$ is the radial velocity 
of each galaxy member and $\langle v \rangle$ is the median of the radial velocity 
of the members. Only 144 of the 230 CGs identified in projection had complete
spectroscopic information. Of these, only 85 CGs passed the velocity filtering
\footnote{Catalogue available at VizieR - cat. J/MNRAS/426/296}.

We here used the sample of filtered 2MCGs selecting the 78 CGs with 
a median group velocity greater than $3000 \, \rm km \, s^{-1}$ 
to avoid introducing effects of peculiar motions 
(see \citealt{diaz12} for further descriptions). We also restricted our sample to 
the area covered by the 2M++ catalogue (see next section for 2M++ description).
The final sample comprises 63 CGs. The CG IDs of this sample are quoted in 
Table~\ref{v2MCG}.
The sky coverage of these groups are shown 
as\textup{ \emph{\textup{filled circles}}} in Fig.~\ref{f1}. The distributions of different 
properties of CGs are shown as \emph{\textup{grey shaded histograms}} 
in Fig.~\ref{f2}.
In this figure we show the distribution of the number of galaxy members in a range of 
three magnitudes from the brightest, 
median radial velocity ($v_{cm}$), 
difference in K-band absolute magnitude between the brightest and second brightest 
galaxies in the group ($K_2-K_1$), 
the K-band apparent magnitude of the brightest galaxy member ($K_{brightest}$), 
the angular diameter of the smallest circle that encloses the galaxy members ($\Theta_G$), 
projected group radius of the smallest circle at 
the distance of the group centre ($R_p$),
and the K-band group surface brightness ($\mu_K$).
We also show the radial velocity dispersion ($\sigma_v$) of 
compact groups computed using the gapper estimator described by \cite{beers90}; 
the 3D group virial radius computed as $R_{vir} = {\pi \over 2 } \, {2 \over \left \langle 1/d_{ij} 
\right\rangle} $, given the inter-galaxy projected separations $d_{ij}$ (see eq. [10--23] of 
\citealp{BT87}); the dimensionless crossing time computed as 
\begin{equation*}
H_0 \, t_{\rm cr} = H_0 \, \frac{\langle d_{ij}^{3D}\rangle}{\sigma_{3D}} =
\frac{100 \, {\rm h} }{ \sqrt{3} \, \sigma_v} \frac{ \pi \, \langle
  d_{ij}\rangle}{2}  \ , 
\label{tcross}
\end{equation*}
where $\langle d_{ij}\rangle$ is the median of
the inter-galaxy projected separations in $\rm h^{-1} \,\rm Mpc$; and 
the virial masses that are obtained by applying the virial theorem \citep{limber60},\begin{equation}
 {\cal M}_{vir} = {3   \, \sigma_v^2 \, R_{vir} \over G} \ . \label{mvir} \end{equation}
Finally, we show the mass-to-light ratio where the group luminosity ($L_K$) is computed by adding 
the individual galaxy luminosities obtained by using the K-correction given by 
\cite{chilinga10}. 
In the first column of Table~\ref{tabcg} in 
the Appendix, we quote the median of these distributions and their semi-interquartile ranges.
\begin{figure}
\begin{center}
\includegraphics[width=8cm]{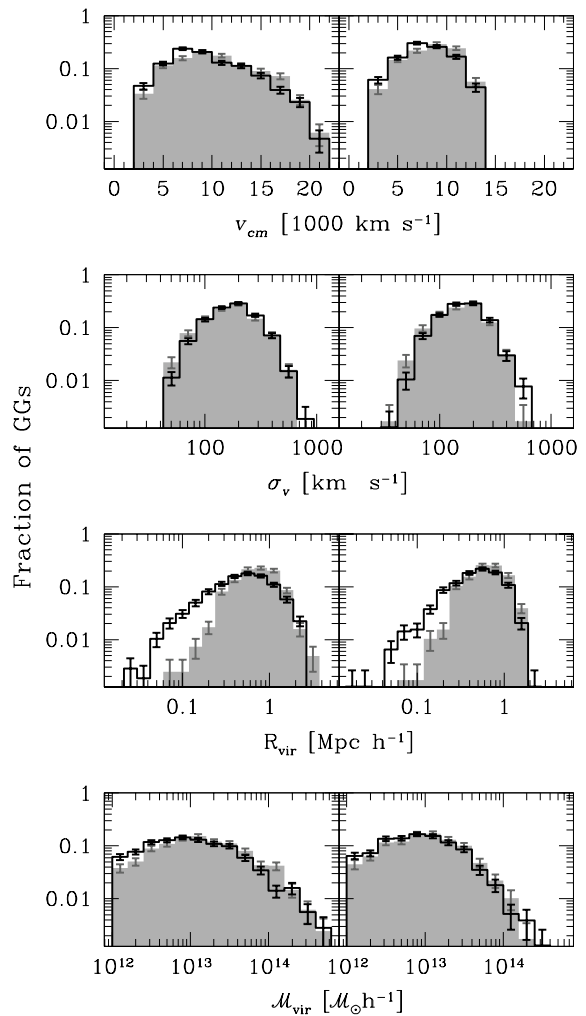}
\caption{Properties of galaxy groups. 
The distributions correspond to
mean radial velocities (first row panels), radial velocity dispersions
(second row panels), 3D virial radii (third row panels) and virial masses 
(fourth row panels) of GGs. Error bars correspond to Poisson errors.
Shaded histograms show the distributions of properties of observational 
galaxy groups 
identified in the 2M++, while empty histograms show the same properties 
for mock galaxy groups identified in a 2M++ mock catalogue.
The left column corresponds to the complete sample of galaxy groups identified by 
the FoF algorithm with a contour overdensity contrast of 433. The right column 
corresponds to the sample of galaxy groups with a radial velocity restricted to span 
the same range as compact groups and excluding the galaxy groups that can be considered
as already included in the sample of compact groups. In this column, 
the mock GGs have been restricted to those that have more than four members after applying 
a blending criterion to account for the size of the particles 
(see Sect.~\ref{mockCG_GG}).  
} 
\label{f3}
\end{center}
\end{figure}

\subsection{Galaxy group sample}
\label{fof} 
To characterise the location of CGs in the local Universe we used as tracers a sample of 
galaxy groups (GGs) identified in the 2M++ 
galaxy redshift catalogue \citep{lavaux11}\footnote{Catalogue available at VizieR - 
cat. J/MNRAS/416/2840}.
This catalogue is based on the 2MASS photometric catalogue for target selection but
with redshift information extracted from the NYU-VAGC for the 
Sloan Digital Sky Survey Data Release Seventh (SDSS-DR7, \citealt{dr7}), 
the Six degree Field Data Release Three (6dF-DR3, \citealt{jones09}), 
and the Two Mass Redshift Survey (2MRS, \citealt{2mrs}).

Since the original 2M++ catalogue has a different magnitude coverage, we concentrated
on the regions where $K_{2M++}\le12.5$ covered by the
SDSS-DR7 or 6dF-DR3, where $K_{2M++}$ is the 2MASS $K_s$ apparent magnitude corrected by 
galactic extinction, galaxy evolutionary effects, and aperture corrections
(see Sect. 2.2 of \citealt{lavaux11} for a complete description of the $K_{2M++}$ magnitude).
We also restricted the sample to galaxies outside the galactic plane ($|b|> 20$).
The resulting galaxy sample comprises 52982 galaxies. 
Their angular distribution in the sky is shown as \emph{\textup{grey points}} in Fig.~\ref{f1}.

The group identification was performed using a friends-of-friends (FoF) algorithm 
similar to that developed by \cite{huchra82} to identify galaxy systems in redshift 
space in a flux-limited catalogue. The algorithm links galaxies that share common neighbours,
that is, pairs of galaxies with projected separations smaller than $D_0$ and radial
velocity differences smaller than $V_0$.
Following the prescriptions of \cite{fof14}, we used a radial linking length of 
$V_0=130\ \rm km \, s^{-1}$ and a transversal linking length, $D_0$, defined by a contour overdensity 
contrast of $\delta\rho/\rho=433$ (see Eq. 4 in \citealt{huchra82}). 
This value of $\delta\rho/\rho$ is adopted
since it is expected that galaxies are more concentrated than dark matter 
\citep{eke04,berlind06}; therefore we should use a higher density contrast than 
that usually adopted in dark matter simulations, between 150-200 
(see Appendix B of \citealt{fof14} for details). 
 
Because of the flux limit of the galaxy sample, both linking lengths
have to be weighted by a factor to take into account the variation 
of the sampling of the luminosity function produced by the different distances 
of the groups to the observers (see Eq. 3 of \citealt{huchra82}). 
This factor is calculated using a Schechter fit of the galaxy luminosity function 
computed for the 2M++ galaxies by \cite{lavaux11}: $M^* - 5 \log{(\rm h )}= -23.43$, $\alpha = -1.03$, 
and $n^* = 0.0085\, \rm h^3 \, Mpc^{-3}$. 
We computed the group physical properties: mean group radial velocity, 
radial velocity dispersions, 3D virial radii and virial masses 
(see Eq.~\ref{mvir} in the previous section and Sect.~5 of \cite{merchan02} 
for further descriptions).

We only selected groups with four or more members, mean group radial velocities
greater than $3000 \, \rm km \, s^{-1}$, and 
virial masses greater than $10^{12} \cal{M}_\odot \, \rm h^{-1}$. 
The lower limit in mass is imposed to avoid galaxy groups with unreliable mass
estimates given the errors in their velocity dispersions.
  
The final GG catalogue comprises 813 objects. 
The distribution of GG properties is shown as shaded histograms 
in the left column of Fig.~\ref{f3}. 
The sample has median radial velocity of $9829 \rm \, km \, s^{-1}$, 
a median radial velocity dispersion of $172 \rm \, km \, s^{-1}$, 
a median virial radius of $0.71 \rm \, Mpc \, \rm h^{-1}$, 
and median virial mass of $1.26 \times 10^{13}\cal{M}_\odot \, \rm h^{-1} $.
This catalogue of galaxy groups is provided in Appendix~\ref{catalogue}.

The FoF algorithm might identify some of the CGs as well as normal groups, and they might also be included in our sample of galaxy groups. 
We therefore examined
the sample of galaxy groups to exclude these groups. 
First, we selected only the galaxies in GGs in a range of three magnitudes 
from its brightest member. Then, to determine whether a GG is also a CG, 
we imposed that at least 75\% of the CG 
members have to be included in the GG, the number of members in the GG in a range
of three magnitudes from the brightest has to be lower than twice the number 
of members in the CG, and only one of those GG member can lie outside the isolation ring 
($3\Theta_G$).
Based on this analysis, we discarded 12 GGs that are already included in the sample of CGs.
The group-IDs of the discarded GGs are 
11, 254, 308, 350, 357, 490, 503, 592, 642, 693, 741, 
and 775  (see first column of Tables~\ref{at1} and~\ref{at2} for references;
in Table~\ref{v2MCG} we flagged the CGs that matched these GGs).

It can be seen from Fig.~\ref{f2} and the left column of Fig.~\ref{f3} that CGs and GGs 
span different ranges of radial velocities. This difference is due to the brightest galaxy 
flux limit criterion imposed to identify CGs. Therefore, we also restricted the sample of GGs to 
those with radial velocities lower than $12500 \rm \, km \, s^{-1}$. 
The final sample comprises 583 galaxy groups. Their angular positions are 
shown as \emph{\textup{open squares}} in Fig.~\ref{f1}. 
The distributions of the main properties of the final GG sample are shown in the 
right column of Fig.~\ref{f3}, while the median of their properties are quoted in 
Table~\ref{tabgg}.

\begin{figure}
\begin{center}
\includegraphics[width=\hsize]{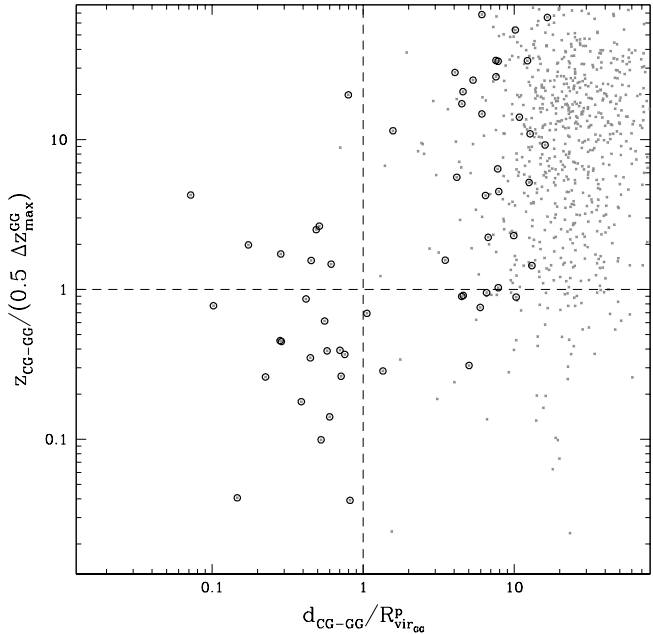}
\caption{Scatter plot of normalised projected
and normalised radial separations between CGs and GGs. 
The grey points show all the separations from each 
CG to all GGs in the sample, while open black circles are the smallest
CG-GG separations only selected in projection. The
dashed lines indicate unity in the normalised
distances. Embedded CGs lie in the lower left corner (see Eq.~\ref{zspace}).
}
\label{fc}
\end{center}
\end{figure}

\section{Location of compact groups in reference to galaxy groups}
\label{place}

As we stated in Sect.~\ref{intro}, nothing prevents CGs from being the core of normal groups 
or smaller substructures in loose groups or filaments. 
In this section we investigate how often this occurs in the local Universe. 
We analysed the location of CGs with 
respect to the location of GGs by classifying CGs into two subsamples: those located inside and those outside GGs (or isolated CGs). 
 
Since we know the positions of CGs in redshift-space (2D+$\frac{1}{2}$), 
the distortions in the line of sight of groups prevent studying 
3D-distances between CGs and GGs. Therefore,
we considered that a CG lies inside a given GG if the projected distance 
between the CG centre and the GG centre at the distance of the GG centre
($d_{CG-GG}$) is smaller than the projected virial radius of the GG 
($R^p_{vir_{GG}}=2\, R_{vir}/\pi$), 
and also the distance in the radial direction among centres ($z_{CG-GG}$) is smaller than 
half the maximum line-of-sight separation among the galaxies identified 
as members of the GG ($\Delta z^{GG}_{max}$), that is, CGs lie inside GGs if 
\begin{equation}
\label{zspace}
{d_{CG-GG} \over R^p_{vir_{GG}}} \le 1 \ \ \ {\rm and} \ \ \ {z_{CG-GG} \over 0.5\, \Delta z^{GG}_{max}} \le 1 
,\end{equation}
where the centre of CGs and GGs in projected and radial directions are  
the barycentre of each system\footnote{For CGs, 
the projected centre is usually assumed 
to be the centre of the smallest circle. However, we adopted the barycentre 
to match the procedure used for GGs.}. 
The scatter plot in Fig.~\ref{fc} shows the distribution of all CG-GG normalised separations in the
projected vs radial directions (grey points) as well as the distribution of CG-GG
smallest normalised separations in the plane of sky (open black circles). 
The lower left region defined by the dashed lines in the scatter plot is the one we used to 
define embedded systems (Eq.~\ref{zspace}). As can be seen
from the figure, selecting groups only by restricting the projected direction
(i.e. projected normalised separation smaller than 1) is not enough to define embedded 
systems, since many of these systems are characterised by larger radial separations
(i.e. radial normalised separation greater than 1).
It is worth mentioning that the greatest projected distance we allowed to consider 
inside/outside GGs is a conservative value ($ 1 \, R^p_{vir_{GG}}$) 
since galaxies in groups 
may extend beyond that limit. 

According to our definition, we found that 17 CGs inhabit GGs, that is, 27\% of the sample of CGs.
We recall that in the previous section, 
we have discarded 12 GGs because they matched some of the CGs in our sample. 
These 12 CGs that are also GGs represent 19\% of the sample of CGs, hence, not including
this constraint would have artificially increased the percentage of CGs located inside GGs.

It is interesting to investigate if the global properties of the CGs are 
affected by their different environments. We analysed the distributions of properties of 
CGs embedded in GGs (27\%) and those that 
could not be directly related to a GG (73\%). 
We compared different properties using
two statistical non-parametric tests: 
a Kolmogorov-Smirnov two-sample test and Mann-Whitney U test that measure
the probability that both samples are drawn from the same distribution.
In Table~\ref{test}, we quote the p-values obtained from the two tests
when comparing the properties of embedded and non-embedded CGs. 
Adopting a typical critical value of 0.05, 
we found that both tests indicate statistical differences for
the distributions of group surface brightness and group projected radius
(highlighted in boldface in Table~\ref{test}). The embedded compact groups are typically brighter and smaller than 
their non-embedded counterparts.

\begin{table}[ht]
\centering
\caption{P-values from two statistical two-sample tests to compare 
the properties of compact groups that can be considered embedded or not. 
The first column is the p-value for the Kolmogorov-Smirnov test (KS), while
the second column is the corresponding with the Mann-Whitney U test (UT). 
We have highlighted in boldface when the tests indicate significant differences 
($p-$value$<0.05$).}
\begin{tabular}{lcc}
\hline
\multicolumn{3}{c}{2MCG}\\
\multicolumn{3}{c}{$N_e = 17$ ; $N_{ne}= 46$}\\
\hline
 Property & KS & UT \\
\hline
$v_{cm}$            & 0.64 & 0.96 \\
$K_2-K_1$           & 0.16 & 0.09  \\
$K_{\rm brightest}$ & 0.80 & 0.81 \\
$\mu_K$             & ${\bf 8\times10^{-3}}$ & $ {\bf6\times10^{-3}}$ \\
$R_p$               &  ${\bf3\times10^{-3}}$ &  ${\bf9\times10^{-3}}$ \\
$\sigma_v$          & 0.80 & 0.97 \\
$R_{vir}$           & 0.15 & 0.47 \\
$H_0 \, t_{cr}$     & 0.52 & 0.27 \\
${\cal M}_{vir}$    & 0.73 & 0.69 \\
${\cal M}_{vir}/\rm L_K$ & 0.97 & 0.93 \\
\hline
\multicolumn{3}{c}{SAM}\\
\multicolumn{3}{c}{$N_e = 40$ ; $N_{ne}= 110$}\\
\hline
$v_{cm}$            & {\bf 0.03} & {\bf 0.04} \\
$K_2-K_1$           & {\bf 0.01} & {\bf 0.02} \\
$K_{\rm brightest}$ & 0.27 & 0.22 \\
$\mu_K$             &  ${\bf 7\times10^{-7}}$ & $ {\bf 7\times10^{-8}}$ \\
$R_p$               & ${\bf 5\times10^{-8}}$ &  ${\bf 5\times10^{-8}}$ \\
$\sigma_v$          & 0.42 & 0.22 \\
$R_{vir}$           & ${\bf 4\times10^{-4}}$ &  ${\bf 2\times10^{-4}}$ \\
$H_0 \, t_{cr}$     & ${\bf 2\times10^{-4}}$ &  ${\bf 2\times10^{-5}}$ \\
${\cal M}_{vir}$    & 0.51 & 0.55 \\
${\cal M}_{vir}/\rm L_K$ & 0.61 & 0.60 \\
\hline
\end{tabular}
\parbox{5cm}{\tiny $N_e$: Number of embedded CGs;\\
 $N_{ne}$: Number of non-embedded CGs}
\label{test}
\end{table}

\begin{table*}
\centering
\caption{Percentages of compact groups that lie within the
boundaries of galaxy groups in redshift-space. 
Each column represents a different 
subsample of CGs, while each row represents a different GG subsample 
(see Sect.~\ref{subsamples} for a more detailed description of each subsample). 
Errors were computed with a bootstrap-resampling technique. 
For each GG subsample, 
we have included a second row with the 
expected percentage of embedded compact groups, which is computed
considering the size of the GG subsample as the product of 
the total percentage, quoted in the first row, 
and the percentage of the GG sample that represents the corresponding subsample, quoted in the first column.
\label{perall} 
}
\tabcolsep 3.5pt
\begin{tabular}{c|ccccccccccccccccccc}
& \multicolumn{2}{c}{ \% of GG} & & \multicolumn{2}{c}{ALL CG} && \multicolumn{2}{c}{Dom CG} && \multicolumn{2}{c}{No-Dom CG} && \multicolumn{2}{c}{Real CG} && \multicolumn{2}{c}{CA CG}\\

\cline{2-3} \cline{5-6} \cline{8-9} \cline{11-12} \cline{14-15} \cline{17-18}

& OBS & SAM && OBS & SAM && OBS & SAM && OBS & SAM && OBS & SAM && OBS & SAM \\
ALL GG & 100 & 100 && $27 \pm 5$ & $27\pm3$ && $34\pm8$ & $22\pm4$ && $19\pm6$ & $32\pm5$ && $--$ & $35\pm6$ && $--$ & $17\pm4$ \\

\cline{1-3} \cline{5-6} \cline{8-9} \cline{11-12} \cline{14-15} \cline{17-18}
Low mass GG & 33 & 40 & & $6\pm2$ & $11\pm2$ && $6\pm4$ & $12\pm4$ && $6\pm5$ & $9\pm2$ && $--$ & $15\pm3$ && $--$ & $6\pm3$ & &  \\
&& & & $9\pm2$ & $11\pm 1$ && $11\pm 3$ & $9\pm 2$ && $6\pm 2$ & $13\pm 2$ && $--$ & $14\pm 2$ && $--$ & $7\pm 2$ & & (*) \\

\cline{1-3} \cline{5-6} \cline{8-9} \cline{11-12} \cline{14-15} \cline{17-18}
Int mass GG & 33 & 32 & & $14\pm5$ & $9\pm2$ && $19\pm6$ & $7\pm3$ && $10\pm5$ & $11\pm3$ && $--$ & $12\pm4$ && $--$ & $4\pm2$& &  \\
& & & & $9\pm 2$ & $9\pm 1$ && $11\pm 3$ & $7\pm 1$ && $6\pm 2$ & $10\pm 2$ && $--$ & $11\pm 2$ && $--$ & $5\pm 1$ & & (*)\\

\cline{1-3} \cline{5-6} \cline{8-9} \cline{11-12} \cline{14-15} \cline{17-18}
High mass GG & 34 & 28 & & $6\pm2$ & $7\pm3$ && $9\pm5$ & $3\pm2$ && $3\pm1$ & $12\pm3$ && $--$ & $7\pm4$ && $--$ & $7\pm3$ & & \\
& & & & $9\pm2$ & $8\pm 1$ && $12\pm 3$ & $6\pm 1$ && $6 \pm 2$ & $9\pm 1$ && $--$ & $10\pm 2$ && $--$ & $5\pm 1$ & & (*)\\

\cline{1-3} \cline{5-6} \cline{8-9} \cline{11-12} \cline{14-15} \cline{17-18}
ALL GG+5 & 100 & 100 && $26 \pm 5$ & $22\pm3$ && $31\pm9$ & $17\pm4$ && $20\pm6$ & $24\pm5$ && $--$ & $26\pm6$ && $--$ & $15\pm5$ \\
\cline{1-3} \cline{5-6} \cline{8-9} \cline{11-12} \cline{14-15} \cline{17-18}
Gaussian GG & 50 & 58 && $13\pm4$ & $12\pm2$ && $16\pm6$ & $9\pm3$ && $10\pm5$ & $14\pm4$ && $--$ & $14\pm5$ && $--$ & $10\pm4$ & & \\
&&&& $13\pm 3$ & $13\pm 2$ && $16\pm 5$ & $10\pm 2$ && $10\pm 3$ & $14\pm 3$ && $--$ & $15\pm 3$ && $--$ & $9\pm 3$ & & (*)\\

\cline{1-3} \cline{5-6} \cline{8-9} \cline{11-12} \cline{14-15} \cline{17-18}
Non-Gaussian GG & 12 & 7 && $2\pm1$ & $1\pm1$ && $3\pm3$ & $0\pm1$ && $0\pm1$ & $1\pm1$ && $--$ & $0\pm1$ && $--$ & $1\pm1$ & & \\
&&&& $3\pm 1$ & $2\pm 1$ && $4\pm 1$ & $1\pm 1$ && $2\pm 1$ & $2\pm 1$ && $--$ & $2\pm 1$ && $--$ & $1\pm 1$ & & (*)\\

\cline{1-3} \cline{5-6} \cline{8-9} \cline{11-12} \cline{14-15} \cline{17-18}
Not classified GG & 38 & 35 && $11\pm3$ & $9\pm2$ && $12\pm6$ & $8\pm3$ && $10\pm4$ & $9\pm3$ && $--$ & $12\pm4$ && $--$ & $4\pm3$ & & \\
&&&& $10\pm 2$ & $8\pm 1$ && $12\pm 3$ & $6\pm 1$ && $8\pm 2$ & $8\pm 2$ && $--$ & $9\pm 2$ && $--$ & $5\pm 2$ & & (*)\\
\hline
\end{tabular}
\vskip 0.2cm
\parbox{18cm}{
\small
{\bf Notes.} OBS: Results obtained from the observational samples. 
SAM: Results obtained from the semi-analytical samples. 
 $(\ast)$: Rows with the expected percentages of embedded CGs for each subsample of GGs.
}

\end{table*}

\subsection{Subsamples of groups}
\label{subsamples}

We have also analysed whether different classes of CGs live 
preferentially inside GGs, and also if different properties of GGs 
make them more suitable to host CGs.

First, we split the sample of CGs according to the K-band absolute magnitude gap between 
the first ranked and the second ranked galaxies in the CG ($K_2-K_1$, see Fig.~\ref{f2}). 
CGs whose magnitude gaps between the brightest and the second brightest galaxies 
are greater than or equal to the median of the $K_2-K_1$ distribution are dominated by 
the brightest galaxy of the systems, and therefore we named this subsample dominated CGs. If the magnitude gap is smaller than the median of 
the distribution of differences, the CG is classified as non-dominated.

Another very important piece of information was obtained by analysing
whether the embedded CGs are preferentially located in a particular type of 
GGs. To determine this, we split the sample of GGs according to their virial masses 
into low-, intermediate-, and high-mass groups using the 
$33^{th}$ and the $66^{th}$ percentiles of the mass distribution.

On the other hand, some of the CGs located inside GGs might be associated
with a particular substructure of the host GG. Since the amount of substructure
in GGs can be inversely associated with the level of relaxation of the groups \citep{einasto12},
one possible way to take into account this situation is to quantify the degree of equilibrium
of a galaxy group by means of its galaxy member velocity distribution, that is, its
internal dynamics. If the distribution has a Gaussian shape, then the 
system can be considered as relaxed,
and a non-Gaussian shape of the velocity distribution is a strong indicator 
of the non-equilibrium state of the galaxy system. 
Hence, we split the sample of GGs according to the Gaussianity of the distributions
of the radial velocities of their galaxy members.  
Following \cite{martinez12}, we applied the Anderson-Darling 
goodness-of-fit test to distinguish between GGs with Gaussian or non-Gaussian velocity
distributions, which can be understood as an indicator of the relaxation of the systems 
and the amount of substructure. For each group, the radial velocity of their 
members and the velocity dispersion of the group was used to compute a parameter, $\alpha$ 
(see Eqs.~7, 8, and 17 of \citealt{hou09}), to distinguish between Gaussian and 
non-Gaussian groups. 
However, according to \cite{hou09}, this classification is statistically reliable 
only when the GG has five or more members, therefore, 
we only applied this test to the 329 GGs 
with $N_{member} \ge 5$. This subsample may include GGs that cannot 
be classified as Gaussian nor non-Gaussian. 

To summarise, the subsamples of GGs are defined as follows: 
\begin{itemize}
\item \emph{\textup{Low-mass}} GGs:  ${\cal M}_{vir} \le 6\times 10^{12} {\cal M}_\odot \, \rm h^{-1}$.
\item \emph{\textup{I}\textup{ntermediate-mass GGs}}: ${6\times 10^{12} {\cal M}_\odot \, \rm h^{-1} < 
\cal M}_{vir}<  1.48\times 10^{13} {\cal M}_\odot \, \rm h^{-1}$. 
\item \emph{\textup{High-mass}} GGs: ${\cal M}_{vir} \ge 1.48\times 10^{13} {\cal M}_\odot \, \rm h^{-1}$. 
\item \emph{\textup{Gaussian}} GGs: GGs with a Gaussian distribution of 
the radial velocity of their galaxy members, i.e., $N\ge 5$ and $\alpha \ge 0.5$ 
\item \emph{\textup{Non-Gaussian}} GGs: GGs with a non-Gaussian distribution 
of the radial velocity of  their galaxy members, i.e., $N\ge 5$ and $\alpha < 0.1$  
\item \emph{\textup{Not classified} (NC)} GGs: GGs that cannot be classified as 
Gaussian or non-Gaussian, i.e., $N\ge5$ and $0.1 \le \alpha < 0.5$ 
\end{itemize} 

The number of groups belonging to each GG subsample and the median of their 
properties are quoted in Table~\ref{tabgg} in the Appendix.

In Table~\ref{perall}, we quote the percentage of CGs that lie inside a GG. 
Each column corresponds to a subsample of CGs, 
while each row corresponds to a subsample of GGs. Errors are computed 
as the mean standard deviation of the percentages obtained from 100 bootstrap 
resamplings of the compact group subsample. 

The column labelled ``ALL CG'' of this table shows that $(27\pm5)$\% of the CGs are located inside GGs (first row). 
We first analysed the preference of the embedded CGs towards particular GG hosts
divided according to the virial
masses of the GGs. The percentage of CGs that inhabit different GG 
subsamples are quoted in the second, third, and fourth rows. 
We found that $(6\pm2)$\%, $(14\pm5)$\% and $(6\pm2)$\% of the CGs inhabit the 
low-, intermediate-, and high-mass GGs, respectively. We also included for each GG 
subsample the percentage of embedded CGs that would be expected if
the distribution of embedded CGs were to depend only on the size of the GG subsamples. This 
expected percentage is computed as the product 
between the total percentage of embedded CGs and the percentage of each of the GG 
subsamples, quoted in the first column of Table~\ref{perall}. 
Errors for the expected values are computed via error propagation. 
The comparison between the measured and expected values shows 
that there is no preference (within 1 sigma significance level) of embedded CGs 
to inhabit any particular GG mass range subsample.

Secondly, we analysed the preference of the embedded CGs for a particular GG subsample 
split according to the Gaussianity of the distribution of the radial velocities 
of their members.
We found that $(26\pm5)$\% of the CGs lie inside the GGs with five or more members 
that are used to split the GGs according to their dynamical state
(see fifth row of the column labelled ``ALL CG'' in Table~\ref{perall}). 
By analysing the frequency of embedded CGs in Gaussian, non-Gaussian, and not-classified GGs 
and comparing these values with the expected ones, we found that there are no 
statistical indications that CGs prefer inhabiting GGs with any particular dynamical state.

Table~\ref{perall} shows that about one third$ $ of the dominated CGs lies inside GGs, 
while only about one fifth$ $ of the non-dominated CGs are located in 
GGs.
The comparison of the percentages of dominated and non-dominated CGs that inhabit
different subsamples of GGs and the expected value shows no preferences
of dominated and non-dominated embedded CGs to be located in GGs in a particular virial mass range or with a particular dynamical state.

\section{Comparison with semi-analytical samples}
\label{sam}
We used a semi-analytic galaxy catalogue to perform the same analyses 
as were developed in the previous sections, with the aim of testing the current
semi-analytic models of galaxy formation, and also to take advantage of
the 3D information that we can access in the models and not in observations. 

\subsection{Mock catalogue}
We adopted the mock all-sky galaxy light cone built by 
\cite{henriques12}, which was constructed by replicating the Millennium I simulation box 
($500 \, \rm Mpc \, \rm h^{-1}$) and taking into account the evolution of structures 
by using the different outputs of the simulation at previous cosmological 
times\footnote{Galaxy mock light cone available as
table wmap1.BC03\_AllSky\_001 at http://www.mpa-garching.mpg.de/millennium/}.
The synthetic galaxies in this light cone were constructed using the 
semi-analytical model of galaxy formation developed by \cite{guo11}. 
This mock catalogue provides the observer frame apparent magnitudes of galaxies
in nine different bands, including the $K_s$ band, 
which we adopted in this work to perform the comparison with observations. 
We could have used the publicly available data from the Millennium II simulation combined 
with the same semi-analytical model, since it has a better resolution in mass. 
However, in a previous work \citep{diaz10}, it has been shown that 
compact groups are dependent on the photometric band in which they are identified: 
a compact group in the K band is not necessarily a compact group in the r band. 
Since we are interested in comparing the results with those obtained from the 2MASS catalogue, we 
chose to work with the sample of \cite{henriques12} that has the K-band magnitudes available. 
It has also been shown that using the Millennium II or the Millennium I produces 
compact groups with very similar properties \citep{diaz12}, 
the differences between the two simulations occur in a mass range that 
is beyond the scope of this work.
We converted the apparent $K_s\rm(AB)$ band available for the mock galaxies 
from the AB system to the Vega system to match the 2MASS magnitudes: 
$K_s {\rm(Vega)}=K_s\rm{(AB)}-1.85$ \citep{blanton05,vega_ab12}.
We adopted an apparent magnitude limit of $K_s=13.57$. 
The number density of this particular mock galaxy catalogue reproduces 
 the number density observed in the 2MASS catalogue
remarkably well.  

\subsection{Compact and galaxy group samples}
\label{mockCG_GG}
We identified mock CGs in the galaxy light cone using the same criteria 
as described in Sect.~\ref{compact}, and 
following \cite{diaz10}, we also included the blending of galaxies in projection
on the plane of the mock sky to take
into account the fact that galaxies in the mock catalogue are point-sized 
particles. The main effect that the blending of galaxies has on the CG catalogue is in the 
number of members that we are able to detect, which is important since one of the 
criteria for identifying CGs is indeed the membership (population).
In that previous work, the authors have used the prescriptions of \cite{shen03} 
to compute the half-light radii as a function of the absolute magnitude in the 
r band of each mock galaxy. They blended two galaxies if their angular separation was
smaller than the sum of their angular half-light radii. 
Motivated by some differences between SAMs and observations pointed out in previous works,
mainly regarding the space density and the projected sizes of CGs, we here slightly changed
the criterion to improve the comparison between the observational and mock samples of CGs. 
First, we split the galaxies into ellipticals and non-ellipticals. 
Following \cite{bertone07}, we
used the ratio between the bulge mass and the total stellar mass provided by 
the semi-analytical model as a proxy for the morphology of our
mock galaxies. We classified as 
ellipticals the galaxies with more than 70 per cent of their stars in the bulge, the remaining galaxies were classified as
non-ellipticals. Then, 
we used the prescriptions of \cite{lange15} to compute the half-light radius 
of each mock galaxy in the K band as a function of the stellar mass of each mock galaxy 
(see Eq. 2 for non-ellipticals and Eq. 3 for ellipticals in that work).  
Finally, we considered two galaxies as blended if the 
angular separation between the two galaxies is smaller than one
and a half times the sum of their 
angular half-light radii.

We identified 380 CGs within a solid angle of $4\pi$. 
Then, we restricted  the sample to the area covered by the 2M++ catalogue 
and excluded the CGs with radial velocities lower than $3000 \ km \, s^{-1}$, 
as we did with observations. We also discarded the mock CGs that had more 
than six members to match the observations (top left panel
of Fig.~\ref{f2}). The final sample comprises 150 mock CGs.
The distributions of CG properties are shown as \emph{\textup{empty histograms}} in 
Fig.~\ref{f2}. 
To compare the properties of observable and mock CGs,~ 
the Kolmogorov-Smirnov and Man-Whitney U-test probabilities are quoted in Table 3. 
In general, the properties of observed 
CGs are well recovered in the mock catalogue, although in the semi-analytic CGs 
we found an excess of CGs that were dominated by a bright galaxy (high $K_2 - K_1$) and had a lower group virial radius.
The differences in these distributions 
are not as significant as was previously reported 
\citep{diaz10,diaz12}. 
On one hand, we checked that discarding the mock CGs with a multiplicity higher 
than the observed multiplicity produced a mock CG sample more similar to the observations 
in all the other properties. The absence of high-multiplicity CGs in the observed 2MCG 
sample might only be a limitation imposed by the lack of redshift 
measurements for all the galaxies in the 2MASS catalogue.
On the other hand,
the stronger criterion for blending galaxies than the one used in previous
works can better  account for the observational effect.

\begin{table}[ht]
\centering
\caption{P-values from different statistical two-sample tests to compare 
the properties of compact groups obtained from observations and from a mock galaxy catalogue. The first column is the p-value for the Kolmogorov-Smirnov test (KS), while
the second column is the corresponding value obtained with the Mann-Whitney U test (UT).}
\begin{tabular}{lcc}
\hline
\hline
 Property & KS & UT \\
\hline
$v_{cm}$            & 0.23 & 0.18 \\
$K_2-K_1$           & ${\bf 9\times10^{-3}}$ & ${\bf 0.01}$  \\
$K_{\rm brightest}$ & 0.64 & 0.89 \\
$\mu_K$             & 0.51 & 0.42 \\
$R_p$               & 0.60 & 0.94 \\
$\sigma_v$          & 0.40 & 0.31 \\
$R_{vir}$           & 0.14 & ${\bf 0.02}$  \\
$H_0 \, t_{cr}$     & 0.14 & 0.08 \\
${\cal M}_{vir}$    & 0.40 & 0.89 \\
${\cal M}_{vir}/\rm L_K$  & 0.65 & 0.68 \\
\hline
\end{tabular}
\label{obs_sam}
\end{table}

The GG sample was identified with the FoF algorithm described 
in Sect.~\ref{fof} applied on the mock galaxy catalogue with the same
sky coverage as the 2M++ sample, 
and having galaxies brighter than $K_{2M++}=12.5$, where 
$K_{2M++}$ is the apparent magnitude of the mock catalogue corrected
in a similar way as in the original 2M++ catalogue, but in this case only corrected
by k-correction and galaxy evolutionary effects: 
$K_{2M++}= K_s+ 1.16 \, (2.1 \, z +0.8 \, z )$ 
\citep{lavaux11}.

We identified 1065 GGs with a contour overdensity contrast of $433$, 
virial masses higher than $10^{12} {\cal M}_\odot \, \rm h^{-1}$, mean radial velocities 
greater than $3000\, \rm km \, s^{-1}$, and with four or more members. 
The property distributions
are shown as empty histograms in the left column of Fig.~\ref{f3}. 
In comparison with the observational 
sample of GGs, the mock GGs tend to have smaller virial radii, and hence, 
the virial mass distribution is slightly shifted towards lower virial masses. 
We then recomputed the membership of each group after applying the blending 
criterion described at the beginning of this section and discarded groups with fewer than four galaxies after blending. Finally, following a similar procedure as the one performed in the observational sample, we
discarded 32 mock GGs that are also CGs and restricted the GG sample to those with 
radial velocities lower than $12500 \, \rm km \, s^{-1}$. 
The final sample comprises 770 mock GGs. The property distributions
are shown 
as empty histograms in the right column of Fig.~\ref{f3}, while 
the median of their properties are quoted in Table~\ref{tabgg}. The statistical
comparison between the mock and observational GGs performed with KS and Mann-Whitney U test 
confirms differences in the virial radii: the mock GGs are smaller than the observational sample.
Nevertheless, the comparison between mock and observational results 
in respect to the analysis of the location of CGs with respect to GGs is not affected by the 
sizes of GGs since we used distances normalised to the virial radii to avoid introducing
a dependence on the sizes of the groups (Eq.~\ref{zspace}).

\subsection{Location of compact groups in redshift space}
We examined the distribution of mock CGs in relation to mock GGs by analysing 
the percentages of CGs that are located in GGs following the criterion 
specified in Eq.\ref{zspace}.

We split the samples of CGs and GGs into the different subsamples defined in 
Sect.~\ref{subsamples}, but we also introduced a new classification for CGs by using 
the 3D information (in real space) available in the simulation. Following \cite{diaz10}, 
we split the sample of CGs into physically dense CGs (Reals) and chance alignments (CAs). 
Reals and CAs were defined based on the physical separations between the 
four closest galaxies. 
Real CGs satisfy $s_4 \le 100 \, \rm kpc \, \rm h^{-1}$ or [$s_4 \le 200 \, \rm kpc \, \rm h^{-1}$
 and $S_\parallel/S_\perp \le 2$], where $s_4$ is the largest interparticle separation 
in real space between the four closest galaxies (or the CG itself for quartets), 
$S_\parallel$ is the largest projected separation, 
and $S_\perp$ is the largest separation in the line of sight of the four closest galaxies 
in the CG. Using this definition, we found that 54\% ($p_R$) of the mock CGs are Reals.

The numbers of CGs in each of the subsamples and the median of some of their properties 
are quoted in Table~\ref{tabcg}. The number and the median of the properties of each 
of the GGs are quoted in Table~\ref{tabgg}.

The results of the location of mock CGs compared to mock GGs are quoted in 
Table~\ref{perall}.  
\begin{figure*}
\begin{center}
\includegraphics[width=12cm]{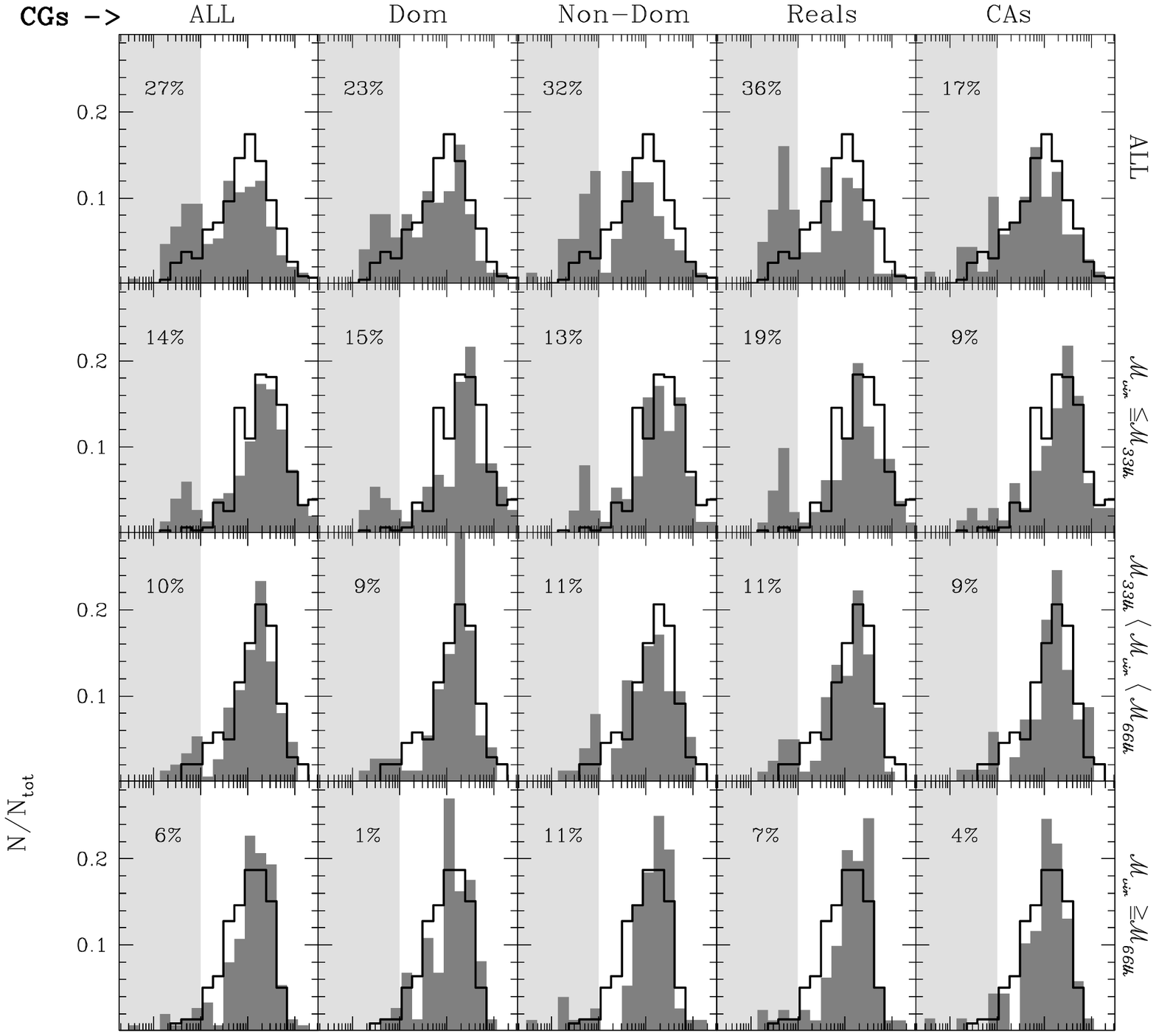}
\vskip -0.055cm
\includegraphics[width=12cm]{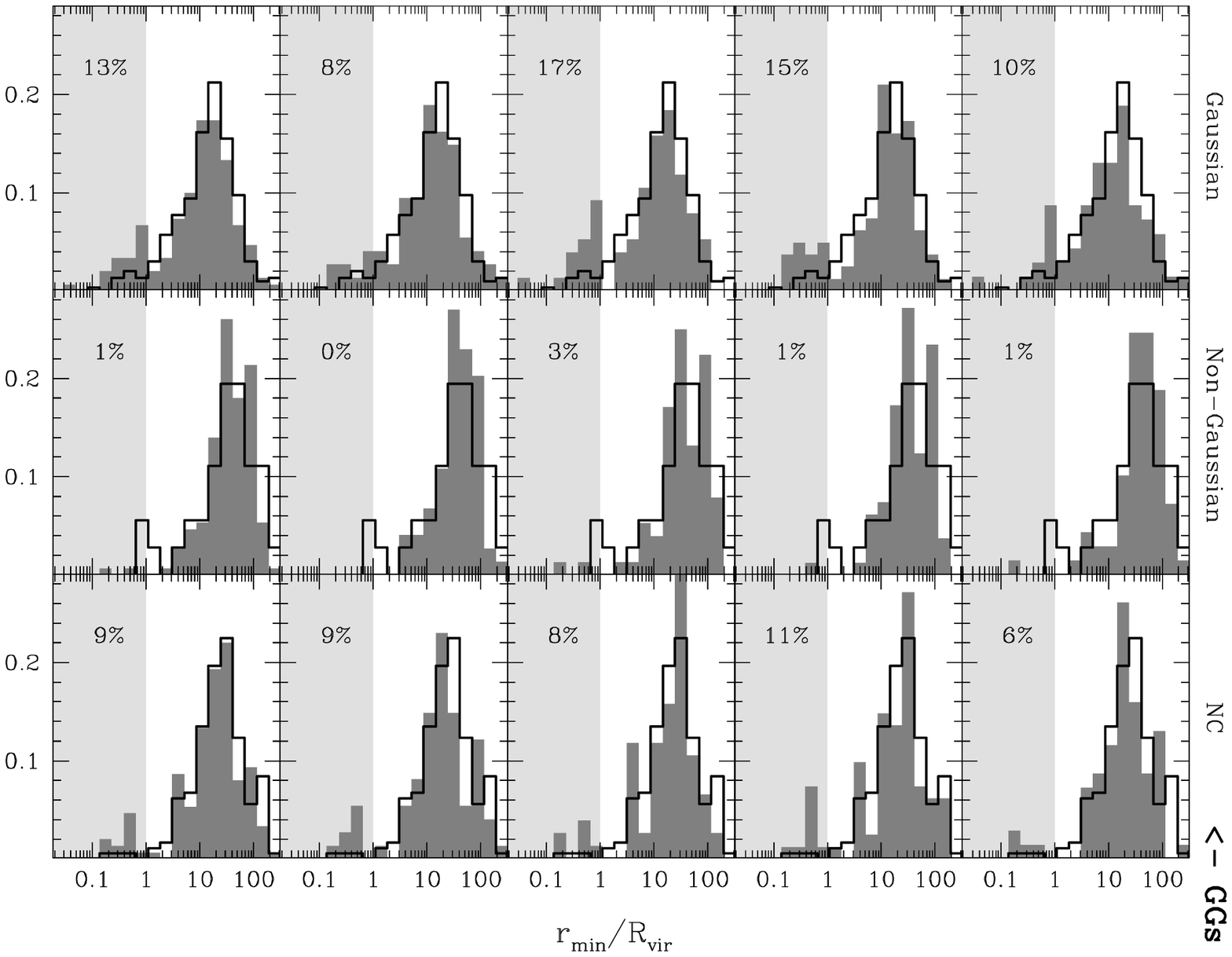}
\caption{Distributions of the 3D smallest normalised separation in real space between
compact and galaxy groups (dark grey histograms). Each column represents a different 
subsample of CGs, each row represents a different GG subsample. 
Each panel also shows the distribution of the 3D smallest separation in real space
among the GGs (black empty histogram).
The light-grey region in each panel determines the inner region of a GG.
Inside that region we quote the percentage of compact groups that are expected to 
lie inside a GG (see Sect.~\ref{subsamples} for a more detailed description of each subsample).
}
\label{f5}
\end{center}
\end{figure*}
The percentage of CGs that lie inside GGs ($p_e=27\%$) 
in the mock catalogue is the same as the percentage found in the observational sample.
This result is very encouraging since it confirms the capability of the current
models of galaxy formation to reproduce the observations of these peculiar environments.
We analysed the distribution of properties of CGs that lie inside and outside GGs. 
The results from the statistical tests are quoted in Table~\ref{test}. 
We found significant differences in the distributions of radial velocities, 
dominance of the brightest galaxy, surface brightnesses, 
projected radii, group virial radii and crossing times, 
with the embedded CGs having brighter surface 
brightnesses, smaller sizes, and shorter crossing times. 
The observations also present similar differences in
surface brightnesses and projected radii. 
This means that observational
and semi-analytical results indicate that embedded CG
are typically smaller in size and brighter than non-embedded CGs.

>From Table~\ref{perall}, analysing the column labelled ``ALL CG'', 
our results show that the mock CGs that inhabit GGs do not have a statistical preference 
towards GGs in any particular mass range, in agreement with the results from observations.
In addition, mock CGs that inhabit GGs with five or more members 
do not have a preference to inhabit GGs with a particular dynamical state 
(Gaussian/non-Gaussian/not classified). 

Some discrepancies can also be observed in the percentages
of dominated and non-dominated CGs that lie inside GGs compared with the
observational results. 
In the mock samples, dominated CGs are less frequent in embedded systems than 
non-dominated CGs (22\% and 32\%), but in observations it is the other way around, 
dominated CGs are more likely to inhabit GGs than non-dominated (34\% vs 19\%). 
However, when analysing the different subsamples of GGs, neither observational or mock dominated/non-dominated CGs show any preferences within 1 sigma significance level.

Finally, the last two columns of Table~\ref{perall} show a piece of information
that we can only access from simulations: there is
a larger percentage of physically dense CGs that lie inside GGs ($p_R^e = 35\%$) than the
corresponding percentage
obtained for CGs considered chance alignments ($p_{CA}^e= 17\%$).
Since the definitions of Reals and CAs can only be made from simulations, 
it is useful to seeking observational clues that can help us select
subsamples of CGs that are dominated by physically dense systems.
By using the previous results, we can predict the percentage of embedded compact groups that 
are Reals. This percentage can be computed as
$p_e^R = p_R \times p_R^e / p_e = 54 \times 35 / 27 = 70\%$. 
On the other hand, performing the same calculation for the non-embedded CGs, 
we found that the percentage of non-embedded CGs that are Reals is $p_{ne}^R=48\%$. 
Therefore, the non-embedded CGs comprise similar percentages of Reals and CAs.
Hence, this result is very useful as a selection criterion in an observational
sample of CGs that are more likely to be Real. Assuming that this prediction can be applied
to observations, by selecting a sample of embedded CGs we can obtain
a sample of CGs where roughly two-thirds of the sample can be considered physically 
dense systems. Applied to 2MCGs, from the 17 2MCGs that are embedded in GGs, we expect 
12 of them to be Real CGs.
 
Finally, taking into account the expected values computed from the relative abundances 
of GG subsamples, we found that Reals and CAs are equally distributed 
within each GG subsample.
 
\subsection{Location of compact groups in real space}
Because real space information is available in the mock catalogue,
in this section we explore the location of CGs around
GGs in the mock catalogue when the redshift distortions are avoided.

For each mock CG, we computed the 3D (in real space) comoving distance 
to its closest GG. The smallest separation between CGs and GGs was computed by 
the Euclidean distance between two points in real space ($r_{\rm min}$). 
For the sake of comparison, we also computed  
the smallest separation among normal groups.
We normalised the distances by the 3D virial radius of the closest GG. 

\emph{\textup{The top left panel}} of Fig.~\ref{f5} shows as \emph{\textup{grey histogram}} 
the distribution of the normalised smallest distances of CGs to GGs. In this figure, 
the \emph{\textup{empty histogram}} is the distribution
of normalised smallest distances of GGs to GGs. The \emph{\textup{light-grey area}} 
represents the region inside unity in the normalised distances. 
The number quoted inside this region is the percentage
of CGs whose smallest distance to a GG is smaller than unity, 
that is, they are positioned inside the virial radius of their closest GG neighbour. 

It can be seen from the \emph{\textup{top left panel}} of this figure that  
27\% of the CGs can be considered as located inside one virial radius of its closest 
GG neighbour in real space. This result agrees with what we found in the previous 
sections.
Hence, this indicates that our criterion to define what is inside
or outside in 
the distorted redshift space is indeed representative of the behaviour in real space. 
Moreover, almost all the percentages quoted
in this figure are quite similar to those obtained in redshift space for all the 
subsamples under analysis.

Finally, the real space information allowed us also to reach a wider understanding
of the location of CGs around GGs in the Universe. 
We found that for most of the subsamples of CGs,
the distribution of distances of CGs to the complete sample of GGs (first row) is clearly bimodal and clearly distinguishable from the distribution of distances of GG-GG, except for the chance alignment CGs, which behave similarly to the GG-GG distribution.
Although this analysis cannot be made in observations, 
hints of this behaviour can be obtained from Fig.~\ref{fc}, which
shows a concentration of points in two different regions of this scatter plot 
(bottom left and top right corner).
  
\section{Summary and conclusions}
We analysed the location of compact groups in the local Universe to determine how important their environment is for the formation scenario of these
systems and for the properties of their galaxy members. 
We used two different approaches, an observational and
a semi-analytical one.

For the observational case, we adopted the compact group catalogue extracted
from the 2MASS catalogue by \cite{diaz12}. This sample has proved to be
a very useful tool since it is one of the largest catalogues with 
spectroscopically confirmed compact groups and homogeneity in the 
selection of the galaxy members that form the groups, it also shows strong
statistical evidence of mergers at the bright end and luminosity segregation.
We adopted as tracer of the large-scale structure of the local Universe a new sample
of galaxy groups identified in the 2M++ galaxy redshift catalogue with a contour
overdensity contrast of 433. 
This galaxy group sample differs from the sample published by \cite{lavaux11} with a
contour overdensity contrast of 80, although the algorithm to identify both is similar 
(Friends-of-Friends).  
Following previous works \citep{eke04,berlind06,fof14,faint14},
we adopted a higher contour overdensity contrast to 
identify galaxy groups in an observational galaxy catalogue.

To perform a comparison with results obtained using mock catalogues, we
adopted the available light cone mock catalogue constructed by \cite{henriques12} using 
the Millennium I simulation plus the semi-analytical model of galaxy formation 
developed by \cite{guo11}. We adapted this mock catalogue to mimic the main 
characteristics of the 2MASS and the 2M++ observational catalogues 
(sky coverage and flux limit)
and proceeded to construct mock compact and galaxy
group samples following the same procedures as used in the observations.

To distinguish between compact groups that lie inside or outside
galaxy systems in redshift space, 
we defined a criterion that takes into account separations in both directions, projected 
and radial. Based on our computations using the observational samples, we
obtained that only $27\%$ of compact groups can be considered to be embedded in overdense
systems, that is, compact groups are more likely to be isolated from galaxy groups. 
This result is lower than the $\sim 50\%$ obtained by
\cite{andernach05,mendel11} and better agrees with \cite{palumbo95,decarvalho05}, who found percentages ranging from $\sim 20\%$ to $\sim 30\%$.
One source for the discrepancy with some of the previous works could
arise from the way our compact group sample was selected: 
only compact groups whose brightest galaxy was fainter than three magnitudes from the magnitude limit of 
the source catalogue were considered. Defining a compact group sample without this restriction includes 
systems that might not fulfil the membership, or the isolation or the 
compactness criteria that might artificially increase the fraction of embedded compact groups.
Another source might be that we carefully selected the galaxy group
 sample to avoid 
considering galaxy groups that are actually compact groups.
We demonstrated that the samples of compact groups
and galaxy groups have an intersection of 12 groups (which represents $19\%$ of the 2MCGs and 
$2\%$ of the galaxy groups). If we had not eliminated these groups
from the 
galaxy group sample, we would have artificially increased the percentage of embedded 
compact groups, 
finding that $46\%$ of the compact group sample are embedded in larger structures, 
similar to the previously reported findings.

We compared the properties of embedded and non-embedded compact groups. 
We found differences in the distributions of surface brightness and projected radius, 
with a tendency of the embedded compact groups to be brighter and smaller 
than the non-embedded compact groups.

We also extended the results by analysing
different subsamples of compact and galaxy groups to determine where those 
embedded compact groups are more likely to be found. 
We observed that embedded compact groups are equally
likely to be found in groups within any virial mass range.
Moreover, using the radial velocity distribution of the group galaxy members to 
characterise the dynamical state of the host galaxy groups, we found no significant evidence 
supporting the idea that embedded compact groups 
are more likely to reside in galaxy groups that have not reached dynamical equilibrium.
Nevertheless, owing to the low number of groups with non-Gaussian distribution
in our sample, these results need to be confirmed with larger samples to increase
the statistical significance of our findings.

When dividing compact groups according to
the dominance of the brightest galaxy (using the magnitude gap between the first and
second brightest galaxies), we also observed that about one third$ $ of the dominated compact groups lies 
inside galaxy groups, while only about one fifth $ $ of the non-dominated compact groups are embedded systems. 
Neither the embedded dominated nor non-dominated compact groups show preferences 
to inhabit galaxy systems within a particular virial mass range. 

After performing a similar analysis on samples of compact and galaxy groups
extracted from a mock galaxy catalogue with semi-analytic information, we observed that
the percentages of embedded mock compact groups in redshift space 
 agree very well with those obtained from the observations. 
Moreover, in both observed and mock compact groups, 
we found that the embedded compact groups tend to be smaller and brighter than the non-embedded 
compact groups.
The semi-analytic catalogue is also able to reproduce the results found in observations
regarding the independence of embedded compact groups to inhabit galaxy groups
of any virial mass and dynamical state.  
On the other hand, we found that mock dominated compact groups are less likely 
to inhabit galaxy groups than observational dominated compact groups, while mock non-dominated 
compact groups are more likely to be found embedded in galaxy groups
 than the observational non-dominated compact groups.
Since the observational sample of compact groups was constructed 
based on the public availability of 
radial velocities of the group members, there might be a bias towards finding 
dominated compact groups (a large galaxy surrounded by small galaxies) 
preferentially inhabiting groups, 
which are the regions that could be more uniformly sampled, and therefore, the fraction of 
embedded dominated compact groups is artificially higher than in the mock catalogue. 
 The observational embedded non-dominated compact groups might
also have been lost because of 
the blending of two similar galaxies located in an overdense environment. 
Although we intended to reproduce the observational
blending of galaxies in the mock catalogue, it might not be enough.  

We also split the compact groups
in the mock catalogue according to a criterion defined by \cite{diaz10} to classify 
these systems into 3D physically dense or chance alignment compact groups. 
>From this analysis, we obtained that the percentage of physically dense compact groups that are 
embedded in galaxy groups is higher than the percentage of embedded chance alignments. 
We found that the embedded compact groups are dominated by physically dense systems ($\sim 70\%$), 
while non-embedded compact groups comprise similar percentages of real and chance alignment 
compact groups. Therefore, this prediction from the semi-analytical sample can
be used as a proxy to obtain a subsample of compact groups dominated by physically dense systems. 
Eigthy-three percent of the chance alignment compact groups
are not embedded in
galaxy groups. This discourages the idea of
chance alignment compact groups being projections inside loose
groups. They might still be projections inside filaments or
the field
, however.

Finally, we used the real space information of the mock catalogue and computed
the distribution of 3D minimum normalised compact group-galaxy group separations for 
each subsample previously analysed in redshift space. 
We found an overall very good agreement with the percentages of embedded compact groups 
obtained by the analysis in redshift space. 
These results clearly support our criterion of defining the smallest
separation in redshift space. Moreover, by analysing the distribution of 
normalised distances, we observed that the shape of the distribution are bimodal for 
several of the subsample combinations, which might be an indication that these are compact groups that may have followed different evolutionary paths depending on the regions they inhabit. 

We conclude that the location of compact groups needs to be carefully taken 
into account when comparing properties of galaxies in compact groups vs galaxies in 
different environments such as normal groups.
Compact groups that are also identified as normal groups 
with the usual group-searching algorithms and compact groups that are embedded in normal systems constitute almost
half of the sample of compact groups. If they are not excluded 
from the galaxy group sample, the comparison of these
samples of galaxy systems 
will be biased by the large intersection among them. Moreover, if
embedded and non-embedded compact groups are not distinguished, the results might be biased as well because
both types of compact groups might be intrinsically different, 
showing different properties, such as the smaller sizes of embedded compact groups, 
which might be related with different formation scenarios.

As a by-product, we release
a new galaxy group catalogue extracted from the 2M++ catalogue that will
be electronically available for the astronomical community (see Appendix~\ref{catalogue}
for details). 

\begin{acknowledgements}
We acknowledge the anonymous referee for insightful comments and suggestions 
that increased the general quality of the paper.
This publication makes use of data products from the Two Micron All Sky Survey, which is a 
joint project of the University of Massachusetts and the Infrared Processing and Analysis 
Center/California Institute of Technology, funded by the National Aeronautics and Space 
Administration and the National Science Foundation.
The Millennium Simulation databases used in this paper and the 
web application providing online access to them were constructed as 
part of the activities of the German Astrophysical Virtual Observatory (GAVO).
We thank Bruno Henriques for allowing public access for his galaxy light cones 
and for kindly answering questions about the data. 
This work has been partially supported by Consejo Nacional de Investigaciones Cient\'\i ficas y 
T\'ecnicas de la Rep\'ublica Argentina (CONICET), Secretar\'\i a de 
Ciencia y Tecnolog\'\i a de la Universidad de C\'ordoba (SeCyT) 
\end{acknowledgements}

\bibliography{refs}
\appendix
\section{Properties of compact groups and galaxy groups}
\label{props}
In Table~\ref{v2MCG} we list the group IDs of compact groups identified by \cite{diaz12}
that have been used in this work. The list includes only the 63 compact groups with 
radial velocities higher than $3000 \, \rm km \, s^{-1}$ and within the sky coverage of the 
2M++ catalogue. The complete catalogue is available at VizieR - cat. J/MNRAS/426/296.

\begin{table}
\begin{center}
\caption{Compact groups from the 2MCG catalogue \label{v2MCG}}
\tabcolsep 3.5pt
\begin{tabular}{ccccccc}
\hline
\multicolumn{7}{c}{group ID}\\
\hline
2 \, \, &5 \, \, & $6^{\ast}$ & 7 \, \, & 8 \, \,& 10 \, & 12 \, \\
14 \,& 15 \,& 16 \,& 17 \,& 18 \,& 19 \,& 20 \, \\
21 \,& $22^{\ast}$& 23 \,& 24 \,& 25 \,& 26 \,& $30^{\ast}$ \\
31 \,& 32 \,& 34 \,& $35^{\ast}$& 36 \,& 37 \,& $38^{\ast}$\\
39 \,& 40 \,& 42 \,& 43 \,& $44^{\ast}$& 45 \,& 46 \,\\
52 \,& 53 \,& $54^{\ast}$& 55 \,& 56 \,& 57 \,& 58 \,\\
59 \,& 60 \,& 61 \,& $62^{\ast}$& $63^{\ast}$& $64^{\ast}$& 65 \,\\
66 \,& 67 \,& 68 \,& 69 \,& 70 \,& $71^{\ast}$& 72 \,\\
74 \,& $75^{\ast}$& 76 \,& 78 \,& 79 \,& 82 \,& 84 \,\\
\end{tabular}
\vskip 0.1cm
\parbox{8cm}{
\small
{\bf Notes.} $\ast$: The 12 compact groups that are also identified as galaxy groups.
}
\end{center}
\end{table}

In Table~\ref{tabcg} we quote the median and semi-interquartile ranges 
of several properties of observational and mock compact groups, 
split into different subsamples.

\begin{table*}
\begin{center}
\caption{Properties of compact groups in different subsamples: median and semi-interquartile ranges \label{tabcg}}
\tabcolsep 3.5pt
\begin{tabular}{c|c|c|c|c|c}
\hline
\multicolumn{6}{c}{2MCG}\\
\hline
Properties & All & Dom  & Non-Dom  & Reals & CAs \\
\hline
Number of CGs & 63 & 32 & 31 & --  & -- \\
$v_{cm} \rm [km \, s^{-1}]$ &
 $  6535 \pm  1604 $ &
$  7375 \pm  2144 $ &
$  5786 \pm  1499 $ & -- & --\\
$K_2 - K_1$ & 
$ 0.9 \pm 0.4 $ & 
$ 1.4 \pm 0.3 $ &
$ 0.6 \pm 0.2 $ & -- & --\\
$K_{brightest}$&
$ 9.9 \pm 0.3 $ & 
$ 10.0 \pm 0.3 $ &
$ 9.8 \pm 0.3 $ &-- & --\\
$\theta_G \, [\rm arcmin]$&
$ 7.3 \pm 3.2 $ & 
$ 6.2 \pm 2.6 $ &
$ 8.8 \pm 3.1 $ &-- & --\\
$R_p \rm [kpc \, \rm h^{-1}]$ & 
$   64 \pm   25 $ &
$   63 \pm   26 $ &
$   74 \pm   25 $ &-- &-- \\
$ \mu_K [\rm mag \, arsec^{-2}]$ &
$ 22.1 \pm  0.7 $ &
$ 22.1 \pm  0.6 $ &
$ 22.2 \pm  0.8 $ & --& --\\
$\sigma_v \rm [km \, s^{-1}]$ & 
$  216 \pm  118 $ &
$  241 \pm  135 $ & 
$  216 \pm   86 $ & -- & --\\
$R_{vir} \rm [kpc \, \rm h^{-1}]$ & 
$   113 \pm   41 $ &
$   113 \pm   41 $ &
$   116 \pm   39 $ &-- &-- \\
$ \rm H_0 \, t_{cr} $&
$ 0.033 \pm 0.028 $ &
$ 0.033 \pm 0.032 $ & 
$ 0.038 \pm 0.023 $ &-- &-- \\
${\cal M}_{vir} [ 10^{12} \, {\cal M}_\odot \, \rm h^{-1}]   $ &
$  5.8 \pm  6.6 $ & 
$  7.4 \pm  7.4 $ & 
$  5.8 \pm  6.6 $ &-- & --\\
${\cal M}_{vir} / \rm L_K \, [{\cal M}_\odot/ L_\odot]$&
$ 38 \pm 33 $ & 
$ 37 \pm 47 $ &
$ 39 \pm 30 $ &-- & --\\
\hline
\multicolumn{6}{c}{Mock CG sample}\\
\hline
Number of CGs &  150 & 75 & 75 & 81  & 69 \\
$v_{cm} \rm [km \, s^{-1}]$ &
$  6847 \pm  1263 $ &
$  7404 \pm  1335 $ & 
$  6569 \pm  1214 $ &
$  6774 \pm  1261 $ &
$  7019 \pm  1296 $ \\
$K_2 - K_1$ & 
$ 1.3 \pm 0.4 $ & 
$ 1.7 \pm 0.3 $ &
$ 0.8 \pm 0.2 $ &
$ 1.2 \pm 0.4 $ &
$ 1.3 \pm 0.5 $ \\
$K_{brightest}$&
$ 9.8 \pm 0.5 $ & 
$ 9.7 \pm 0.5 $ &
$9.9 \pm 0.5 $ &
$ 9.8 \pm 0.4 $ &
$ 9.8 \pm 0.6 $ \\
$\theta_G \, [\rm arcmin]$&
$ 7.3 \pm 2.5 $ & 
$ 7.4 \pm 2.6 $ &
$ 6.7 \pm 2.3 $ &
$ 6.2 \pm 2.1 $ &
$ 7.8 \pm 2.5 $ \\
$R_p \rm [kpc \, \rm h^{-1}]$ &
$   70 \pm   23 $ &
$   75 \pm   23 $ &
$   67 \pm   24 $ & 
$   65 \pm   20 $ &
$   80 \pm   24 $ \\
$ \mu_K [\rm mag \, arsec^{-2}]$ &
$ 22.1 \pm  0.8 $ &
 $ 22.2 \pm  0.9 $ &
$ 22.0 \pm  0.7 $ &
$ 21.8 \pm  0.7 $ &
$ 22.4 \pm  0.7 $ \\ 
$\sigma_v \rm [km \, s^{-1}]$ & 
$  254 \pm   90 $ &
$  232 \pm  92 $ & 
$  255 \pm  97 $ &
$  267 \pm  90 $ &
$  229 \pm   86 $ \\
$R_{vir} \rm [kpc \, \rm h^{-1}]$ &
$  101 \pm   34 $ &
$  107 \pm   33 $ &
$   87 \pm   35 $ & 
$   92 \pm   30 $ &
$   126 \pm   47 $ \\
$ \rm H_0 \, t_{cr} $&
$ 0.030 \pm 0.017 $ &
 $ 0.034 \pm 0.017 $ &
 $ 0.022 \pm 0.014 $ &
$ 0.023 \pm 0.013 $ &
$ 0.036 \pm 0.022 $ \\
${\cal M}_{vir} [ 10^{12} \, {\cal M}_\odot \, \rm h^{-1}]   $ &
$  6.7 \pm  5.5 $ &
$  7.2 \pm  5.3 $ &
$  5.0 \pm  5.6 $ &
$  5.0 \pm  4.7 $ &
$  7.0 \pm  7.0 $ \\
${\cal M}_{vir} / \rm L_K \, [{\cal M}_\odot/ L_\odot]$&
$ 33 \pm 24 $ & 
$ 37 \pm 27 $ &
$ 26 \pm 23 $ &
$ 29 \pm 22 $ &
$ 33 \pm 26 $ \\
\hline
\end{tabular}
\vskip 0.1cm
\end{center}
\end{table*}

In Table~\ref{tabgg} we quote the median and semi-interquartile ranges 
of several properties of observational and mock galaxy groups, 
split into different subsamples.
\begin{table*}
\begin{center}
\caption{Properties of galaxy groups in different subsamples: median and semi-interquartile ranges.\label{tabgg}}
\tabcolsep 3.5pt
\begin{tabular}{c|c|c|c|c|c|c|c}
\hline
\multicolumn{8}{c}{Observational group sample}\\
\hline
Properties & All & Low  & Intermediate  & High & Gaussian & Non-Gaussian & Not classified \\
 &  &  mass  &  mass &  mass &  &  &  \\

\hline
Number of GGs & 583 & 193 & 191 & 199 & 163 & 39 & 127 \\
$v_{cm} \rm [km \, s^{-1}]$ & 
$ 8739 \pm  1866 $ &
$ 7517 \pm  1802 $ &
$ 8331 \pm  1726 $ &
$ 9789\pm  1477 $ & 
$ 8427 \pm  2049 $ &
$ 8855 \pm  2141 $ &
$ 8808 \pm  1692 $ \\
$\sigma_v \rm [km \, s^{-1}]$ & 
$ 157 \pm 50 $ &
$ 94 \pm 20 $ &
$ 155 \pm 27 $ &
 $ 228 \pm 37 $ &
$ 189 \pm 47 $ &
$ 121 \pm 41 $ &
 $ 154 \pm 52 $ \\
$R_{vir} \rm [Mpc \, \rm h^{-1}]$ & 
$ 0.64 \pm 0.22 $ &
$ 0.51 \pm 0.20 $ &
$ 0.57 \pm 0.19 $ &
$ 0.79 \pm 0.21 $ &
$ 0.68 \pm 0.17 $ &
$ 0.76 \pm 0.19$ &
 $ 0.66 \pm 0.24 $ \\
${\cal M}_{vir} [ 10^{12} \, {\cal M}_\odot \, \rm h^{-1}]   $ &
$ 9.7 \pm 7.7 $ & 
$ 3.2 \pm 1.1 $ &
 $ 9.5 \pm 2.1 $ &
$ 26.1 \pm 10.2 $ &
$ 15.0 \pm 9.3 $ &
$ 7.0 \pm 6.8 $ &
$ 9.5 \pm 8.4 $ \\
\hline
\multicolumn{8}{c}{Mock group sample}\\
\hline
Number of GGs & 770 & 309 & 242 & 219 & 297 & 36 & 178 \\ 
$v_{cm} \rm [km \, s^{-1}]$ & 
$ 7850 \pm  1718 $ &
 $ 7365 \pm  1613 $ &
$ 7858 \pm  1655 $ &
$ 8716 \pm  1634 $ &
$ 7766 \pm  1809 $ &
 $ 7612 \pm  1889 $ &
$ 8010 \pm  1722 $ \\
$\sigma_v \rm [km \, s^{-1}]$ &
$ 160 \pm 51 $ &
$ 114 \pm 30 $ &
$ 160 \pm 32 $ &
$ 236 \pm 45 $ &
$ 188 \pm 47 $ &
$ 185 \pm 72 $ &
$ 156 \pm 53 $ \\ 
$R_{vir} \rm [Mpc \, \rm h^{-1}]$ & 
 $ 0.50 \pm 0.21 $ &
$ 0.33 \pm 0.16 $ &
$ 0.50 \pm 0.19 $ &
 $ 0.70 \pm 0.21 $ &
$ 0.49 \pm 0.21 $ &
$ 0.45 \pm 0.19 $ &
$ 0.45 \pm 0.22 $ \\
${\cal M}_{vir}  [ 10^{12} \, {\cal M}_\odot \, \rm h^{-1}]$ &
$ 8.1 \pm 6.4 $ &
$ 3.1 \pm 1.1 $ &
 $ 9.4 \pm 2.0 $ &
$ 25.7 \pm 8.5 $ &
$ 10.2 \pm 8.4$ &
$ 7.8 \pm 12.1 $ &
$ 7.0 \pm 6.5 $ \\
\hline
\end{tabular}
\vskip 0.1cm
\end{center}
\end{table*}

\section{Group catalogue with $\delta\rho/\rho=433$}
\label{catalogue}
In this section we quote the content of the tables of galaxy groups identified
in the 2M++ catalogue using a contour overdensity contrast of 433.
In Table~\ref{at1} we show part of the table containing the 813 
galaxy groups, while Table~\ref{at2} shows also part of the table that quotes 
the information for the 4869 galaxy members.

\begin{table*}[ht]
\centering
\caption{Galaxy groups identified in the 2M++ catalogue \citep{lavaux11}. 
We include here only a few lines. The complete table can be found in electronic format.}
\begin{tabular}{cccccccc}
\hline
 Id  & $N_g$ & RA & DEC & $v_{cm}$ & $\sigma_v$ &   $R_{vir}$      &           ${\cal M}_{vir}$             \\
       &       & [$deg$]  & [$deg$]  & [$km/s$] &  [$km/s$]  & [$\rm Mpc \ \rm h^{-1}$] &  [$10^{14} \, \cal{M_\odot} \, \rm h^{-1}$]\\
\hline
   ~~1  & ~~4 &    324.127 &  -83.894 &   18002.965  &   224.195  &  1.250  &  0.4383 \\
   ~~2  & ~~4 &  ~~~~5.229 &  -81.026 &   16871.429  &   213.153  &  1.464  &  0.4640 \\
   ~~3  & ~~4 &    337.707 &  -80.216 &   11734.202  &   187.328  &  1.208  &  0.2956 \\
   ~~4  & ~~6 &    336.104 &  -80.243 &   11217.477  &   223.655  &  0.515  &  0.1796 \\
   ~~5  & ~~5 &    253.364 &  -79.943 &   12761.345  &   271.959  &  1.104  &  0.5697 \\
   ~~6  &  10 &    278.508 &  -77.019 &  ~~5597.198  &  ~~84.888  &  0.922  &  0.0463 \\
   ~~7  & ~~4 &   ~~74.464 &  -74.851 &  ~~5806.220  &   193.855  &  0.598  &  0.1568 \\
   ~~8  & ~~5 &    101.298 &  -74.279 &  ~~6377.974  &   200.810  &  0.383  &  0.1076 \\
   ~~9  & ~~4 &   ~~94.245 &  -74.193 &   11696.234  &  ~~76.082  &  0.751  &  0.0303 \\
    10  & ~~7 &    101.903 &  -71.665 &  ~~4281.469  &  ~~80.630  &  0.730  &  0.0331 \\
    11  & ~~9 &    304.395 &  -70.813 &  ~~3710.980  &   223.876  &  0.115  &  0.0403 \\
    12  & ~~4 &    285.085 &  -69.761 &   14098.694  &   193.275  &  1.368  &  0.3565 \\
    13  & ~~4 &   ~~93.125 &  -67.870 &   10871.720  &  ~~89.479  &  0.890  &  0.0497 \\ 
    14  & ~~4 &    339.991 &  -66.724 &   12378.471  &   283.765  &  0.441  &  0.2479 \\
    15  & ~~7 &   ~~92.323 &  -65.670 &   10952.375  &   254.096  &  0.893  &  0.4020 \\
    16  & ~~4 &    103.901 &  -65.527 &  ~~8988.492  &   135.257  &  0.495  &  0.0631 \\
    17  & ~~5 &   ~~94.878 &  -65.094 &  ~~8748.338  &   119.096  &  1.161  &  0.1149 \\
    18  & ~~4 &    341.554 &  -65.201 &  ~~3389.487  &   139.634  &  0.429  &  0.0584 \\
    19  &  24 &   ~~95.603 &  -64.903 &  ~~7729.567  &   439.018  &  0.688  &  0.9251 \\
    20  & ~~5 &    101.649 &  -64.124 &  ~~3485.746  &  ~~66.548  &  0.554  &  0.0171 \\
\hline
\end{tabular}
\label{at1}
\parbox{11cm}{
\small
{\bf Notes.} 
Group ID, Number of galaxies linked by the FoF algorithm, RA: Group centre right ascension (J2000), Dec: group centre declination (J2000), $v_{cm}$: mean group radial velocity, $\sigma_v$: radial velocity dispersion, $R_{vir}$: 3D virial radius, ${\cal M}_{vir}$: group virial mass.\\
Groups in this table were identified using a FoF algorithm with a contour overdensity contrast of 433, and having four or more members, mean group radial velocities greater than $3000 \, \rm km \, s^{-1}$, 
and virial masses greater than $10^{12} \cal{M}_\odot \, \rm h^{-1}$.
}

\end{table*}

\begin{table*}[ht]
\centering
\caption{Galaxy members of groups identified in the 2M++ catalogue \citep{lavaux11}. 
We include here only a few lines. The complete table can be found in electronic format. 
}
\begin{tabular}{cccccc}
\hline
 Id  & gid  & RA & DEC & $z$ & $K_{2M++}$ \\
       &       & [$deg$]  & [$deg$]  &     &            \\
\hline
     1  &  1 &    324.29554 &  -83.72072 &  0.05902  &  11.270 \\
     1  &  2 &    324.05187 &  -83.94928 &  0.06031  &  11.480 \\
     1  &  3 &    323.39417 &  -83.96389 &  0.06044  &  12.030 \\
     1  &  4 &    324.76146 &  -83.93817 &  0.06044  &  12.200 \\
     2  &  1 &  ~~~~4.75204 &  -81.02767 &  0.05622  &  12.220 \\
     2  &  2 &  ~~~~4.91983 &  -80.81486 &  0.05669  &  12.400 \\
     2  &  3 &  ~~~~6.20658 &  -81.20600 &  0.05676  &  12.470 \\
     2  &  4 &  ~~~~5.05354 &  -81.05369 &  0.05544  &  12.490 \\
     3  &  1 &    337.08733 &  -80.44439 &  0.03974  &  10.570 \\
     3  &  2 &    338.20896 &  -80.15331 &  0.03934  &  11.250 \\
     3  &  3 &    337.76642 &  -80.22136 &  0.03879  &  11.960 \\
     3  &  4 &    337.75412 &  -80.03800 &  0.03870  &  11.980 \\
     4  &  1 &    336.03079 &  -80.17547 &  0.03793  &  10.780 \\
     4  &  2 &    335.98571 &  -80.22322 &  0.03770  &  11.290 \\
     4  &  3 &    336.06921 &  -80.43606 &  0.03788  &  11.640 \\
     4  &  4 &    335.89050 &  -80.33175 &  0.03789  &  11.990 \\
     4  &  5 &    336.54004 &  -80.10231 &  0.03697  &  12.110 \\
     4  &  6 &    336.11046 &  -80.18550 &  0.03613  &  12.360 \\
     5  &  1 &    253.20029 &  -79.83975 &  0.04344  &  11.190 \\
     5  &  2 &    253.87708 &  -80.09672 &  0.04158  &  11.620 \\
     5  &  3 &    252.97533 &  -80.03006 &  0.04294  &  11.730 \\
     5  &  4 &    253.66292 &  -80.06969 &  0.04202  &  11.920 \\
     5  &  5 &    253.14283 &  -79.68667 &  0.04285  &  11.970 \\
\hline
\end{tabular}
\label{at2}
\parbox{11cm}{
\small
{\bf Notes.} group id, galaxy index, RA: right ascension (J2000), DEC: declination (J2000),
$z$: heliocentric redshift, $K_{2M++}$: $K_s$ apparent magnitude provided in the 2M++ catalogue.
Galaxies within each group are ordered by their apparent magnitudes from brightest to faintest.
}

\end{table*}

\end{document}